# Twin Peak Method for Estimating Tissue Viscoelasticity using Shear Wave Elastography

**Shuvrodeb Adikary**[1], **Matthew W. Urban**[2], **Murthy N. Guddati**[1]*

[1] *North Carolina State University, Raleigh, NC, US.* [2]*Mayo Clinic, Rochester, MN, USA*

* *Corresponding Author. E-mail: mnguddat@ncsu.edu*

## Abstract

Tissue viscoelasticity is becoming an increasingly useful biomarker beyond elasticity and can theoretically be estimated using shear wave elastography (SWE), by inverting the propagation and attenuation characteristics of shear waves. Estimating viscosity is often more difficult than elasticity because attenuation, the main effect of viscosity, leads to poor signal-to-noise ratio of the shear wave motion. In the present work, we provide an alternative to existing methods of viscoelasticity estimation that is robust against noise. The method minimizes the difference between simulated and measured versions of two sets of peaks (twin peaks) in the frequency-wavenumber domain, obtained first by traversing through each frequency and then by traversing through each wavenumber. The slopes and deviation of the twin peaks are sensitive to elasticity and viscosity respectively, leading to the effectiveness of the proposed inversion algorithm for characterizing mechanical properties. This expected effectiveness is confirmed through *in silico* verification, followed by *ex vivo* validation and *in vivo* application, indicating that the proposed approach can be effectively used in accurately estimating viscoelasticity, thus potentially contributing to the development of enhanced biomarkers.

## Introduction

There has been growing interest in estimating tissue viscoelasticity using ultrasound and magnetic resonance imaging given its role as a non-invasive biomarker for many diseases [1–4]. Different elastography techniques such as transient elastography (TE, FibroScan), magnetic resonance elastography (MRE), and ultrasound shear wave elastography (SWE) have been used to characterize disease progression [5–9]. TE primarily measures only elasticity leading researchers to focus on MRE and SWE for viscoelasticity inversion, and despite the accuracy of MRE [6,10–13], the cost associated with MRE is significantly higher than TE or SWE [12,14]. We thus focus on SWE given that it is widely available, inexpensive, and can estimate both elasticity and viscosity.

Many initial efforts using ultrasound-based methods to characterize tissue viscoelasticity were based on evaluating the phase velocity and assessing the dispersion, i.e., the variation of the phase velocity with frequency [15]. The dispersion curves (phase velocity as a function of frequency) can be measured with different signal processing approaches such as the phase gradient and two-dimensional (2D) Fourier transform, as well as some additional recently proposed methods [16–22]; the resulting curves can be fit with different rheological models of viscoelasticity [15,17,20,23–25]. Rouze et al. [26] extended the idea of dispersion matching and utilized the difference between the shear wave speeds from particle displacements, velocities and accelerations to invert for tissue viscoelasticity.

As an alternative to shear wave dispersion, shear wave velocity and attenuation can be used together to characterize the complex-valued viscoelastic modulus [27]. Compared to shear wave velocity, shear wave attenuation can be difficult to estimate because of the need to incorporate information about the shear wave source. Many acoustic radiation force (ARF) push beams can be approximated as a cylindrical source and the attenuation due to the diffraction of the source needs to be accounted for in the estimation process [28]. Budelli et al. [29] proposed a method for reconstructing images of shear wave velocity and attenuation by accounting for these diffraction effects. Nenadic et al. [28] developed the AMUSE method to estimate the phase velocity and attenuation, after incorporating diffraction correction. In subsequent work, Rouze et al. [30] noted the reduced accuracy of the AMUSE method due to truncation of signals and proposed finite window, 2D Fourier transform (fw2DFT) to remove this limitation. Later, Kijanka and Urban [31] proposed the two-point frequency shift approach, building on the initial contribution of Bernard et al. [32], who utilized the ideas of amplitude spectra frequency shift to measure attenuation. Kijanka and Urban [33] further extended this study for better characterization of attenuation using power spectra frequency shift of shear waves measured at two spatial positions. Yazdani et al. [34] also used a modified frequency shift approach to prevent the outlier attenuation values in the presence of noise.

The other related contributions include the use of convolutional neural networks [35] and full wave inversion [36]. Despite the many contributions, viscoelasticity estimation is inherently difficult given that the adverse effects of noise amplify due to the signal decay associated with tissue viscosity. In this paper, we propose a new approach of viscoelastic inversion of SWE data focusing on the signatures that are more robust against noise, i.e. peaks of the response in Fourier space. We first observe that in the presence of viscoelasticity, the peaks of the particle velocity response (in the Fourier domain) can be defined in two different ways: $k(f)$ peaks, which are wavenumbers associated with maximum response while sweeping through frequencies, and $f(k)$ peaks which are correspondingly the frequencies at the maximum response while sweeping through wavenumbers. Importantly, these two peaks diverge, and the extent of the divergence depends on tissue viscosity, indicating that matching both these peaks would lead to inversion of both elasticity and viscosity. We thus call this the twin-peak method (TPM), which is the main subject of the paper.

This paper is organized as follows. After defining the problem setup and theoretical background, we summarize the mathematical rationale behind the TPM method. We then move on to the details of the method by first describing a forward model that takes in viscoelastic parameters as input and computes simulated $f(k)$ and $k(f)$ peaks. The forward model is then used in an iterative inversion framework that translates experimental $f(k)$ and $k(f)$ peaks to estimates of viscoelasticity parameters. Finally, the TPM's accuracy is examined using *in silico* and *ex vivo* experiments, followed by *in vivo* application.

## Methods

### Shear Wave Elastography: Problem Setup

Shear waves are generated using focused acoustic radiation force (ARF), applied by a focused ultrasound beam at a specific region of the tissue (see Figure 1a). The resulting wave is measured as particle velocity in the $z$ direction on the $x-z$ plane. Considering the noise in the signal, the particle velocity is often averaged in the $z$ direction over a small horizontal strip at the center depth of the push, resulting in the particle velocity representation in the $x-t$ domain (see Figure 1b). The objective is then to process the $x-t$ representation of particle velocity to estimate the tissue viscoelasticity.

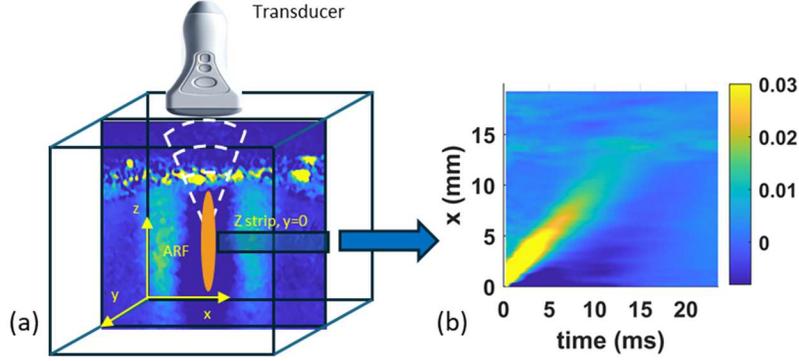

**Figure 1.** (a) Experimental setup of SWE, (b) $x-t$ representation of measured particle velocity response.

In this study, we consider the tissue to be a homogenous bulk, and ARF is essentially a time-dependent body force. The resulting displacement can be obtained by the incompressible elastodynamics equation with the field variable being the three-dimensional (3D) displacement vector. Given the dominance of shear waves and the relatively long shape of the ARF along the $z$ axis, the equation is often simplified into a wave equation with the scalar field variable $u$ representing the displacement in the $z$ direction:

$$\rho \frac{\partial^2 u}{\partial^2 t} - G * \nabla^2 u = f. \qquad (1)$$

$\rho$ is the density, $G$ is the (viscoelastic) relaxation modulus in shear, $\nabla^2$ is the Laplacian, $f$ represents ARF. The * symbol is the convolution operator capturing the memory effects of viscoelasticity. The ARF is characterized by its shape in space $f_s$, multiplied with a rectangular step function in time $f_t$, i.e.,

$$f(x,y,z,t) = f_s(x,y,z) \times f_t(t), \qquad (2)$$

$$f_t = \begin{cases} 1 & 0 \le t \le T, \\ 0 & otherwise, \end{cases} \qquad (3)$$

where $T$ is the duration of the ARF, which is chosen to be 400 microseconds in this study. The measured response $u$ is linked to the unknown viscoelastic modulus $G$. Our goal is to estimate the unknown modulus $G$ from measured $u$ on the SWE along the $x$ axis.

## Basic Idea: Twin-Peak Method (TPM)

The key to our approach is the observation that the $x-t$ particle velocity, examined in the frequency $(f)$ - wavenumber $(k_x)$ domain has a peak with wide spread compared to that from an elastic medium (see Figure 2a, resulting from the data in Figure 1 and the procedure in the Section, *Identifying Peaks*). The spread between peaks is a function of viscosity. This spread, quantified using the full-width at half-maximum [28], was used to estimate viscosity but is prone to noise. We instead quantify the spread by examining two peaks (see Figure 2d), so called $f(k)$ and $k(f)$ peaks, which are distinct and are expected to be robust against noise. The $f(k)$ peaks are obtained from examining $f-k$ transform of the SWE signal for different wavenumbers $k_x$ and finding the corresponding frequency $f$ associated with the maximum absolute value of the response (this can be easily visualized by plotting the absolute value of $f-k$ response after normalizing by the maximum value for each $k_x$ as shown in Figure 2b). The $k(f)$ peaks are similarly obtained by locating the wavenumber $k$ corresponding to maximum response for each $f$ (visualized again with appropriate normalization as shown in Figure 2c). The two peaks are shown together in Figure 2d, which clearly illustrates the spread, indicating that viscoelasticity can be estimated by fitting both peaks. While at the end we utilize standard optimization-based inversion technique to estimate viscoelasticity from these peaks, in the remainder of the section, we provide basic mathematical insight into the method with the help of simplified 1D wave propagation model.

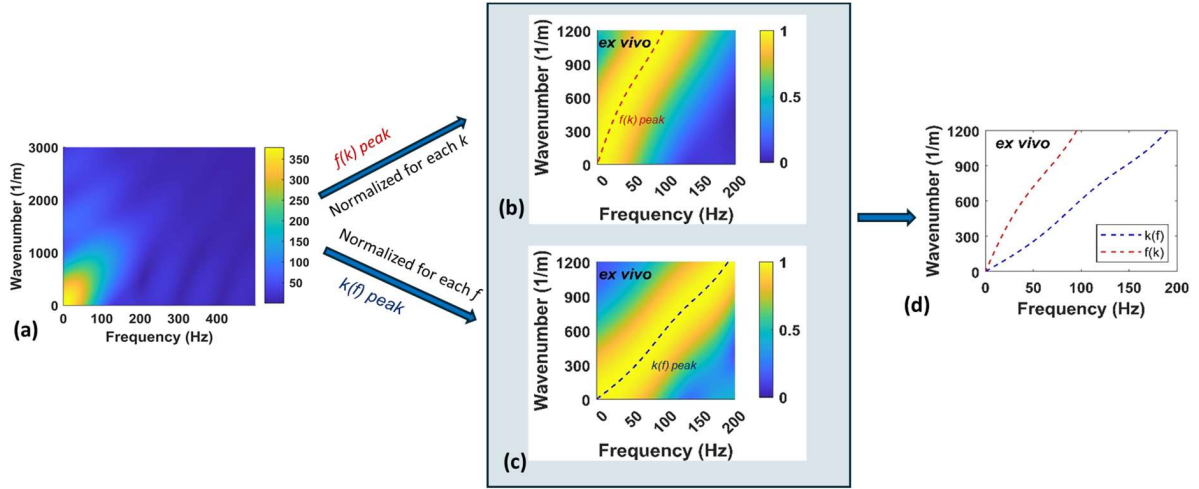

**Figure 2.** (a) Particle velocity in frequency-wavenumber domain, (b) visualization of $f(k)$ peaks by normalizing $f-k$ response for each $k_x$, (c) visualization of $k(f)$ peaks by normalizing $f-k$ response for each $f$, (d) illustration of the two diverging peaks.

Specializing Equation 1 to an infinitely long axisymmetric push leads to a one-dimensional (1D) axisymmetric wave equation written in polar coordinates. Applying $\sqrt{r}$ modulation, i.e. $\hat{u} = \sqrt{r}u$ leads to a standard 1D wave equation that governs the far-field :

$$\rho \frac{\partial^2 \hat{u}}{\partial^2 t} - G * \frac{\partial^2 \hat{u}}{\partial^2 x} = f_t(t)\delta(x), \tag{4}$$

where the ARF shape is assumed to be a Dirac delta $(\delta)$ concentrated at the origin. Note here that the $x$ coordinate coincides with the radial direction and hence used interchangeably. Fourier transforming Equation 4 in $x$ and $t$ results in

$$-\rho \omega^2 \hat{U} + \widehat{G} k_x^2 \hat{U} = F(\omega), \tag{5}$$

where $\omega$ and $k_x$ are the temporal and spatial frequency respectively, and $\hat{U} = \hat{U}(k_x, \omega)$ and $F = F(k_x, \omega)$ are the Fourier transforms of the displacement and ARF respectively, $\widehat{G}$ is thus the frequency-dependent complex shear modulus, which is the Fourier transform of the relaxation modulus. Focusing on the simpler Kelvin-Voigt rheological model, $\widehat{G}$ is given by,

$$\widehat{G} = G_0(1 + i\tau\omega) = \rho c_s^2 (1 + i\tau\omega), \tag{6}$$

where $G_0$ is the storage modulus, $c_s$ is the shear wave velocity, and $\tau$ is the relaxation time (the ratio between viscosity and elastic modulus). Further, idealizing ARF to be an impulse results in $F=1$, leading to the frequency-wavenumber representation of the displacement:

$$\hat{U}(k_x,\omega) = \frac{1}{\widehat{G}k_x^2 - \rho\omega^2} = \frac{1}{\rho\left(c_s^2 k_x^2(1+i\tau\omega) - \rho\omega^2\right)}. \tag{7}$$

The particle velocity, being the time derivative of displacement, is given in frequency domain,

$$\hat{V}(k_x,\omega) = i\omega\hat{U} = \frac{i\omega}{\rho\left(c_s^2 k_x^2(1+i\tau\omega) - \rho\omega^2\right)}. \tag{8}$$

Peaks of this response can be defined in two ways: (a) for each wavenumber $k_x$, obtain the frequency at which the amplitude of the response is maximum, leading to $f(k)$ peak, and (b) obtain $k(f)$ peak by traversing along $f$, and locating $k_x$ associated with maximum response:

$$f(k) \text{ peak:} \quad \omega(k_x) = \arg\max_\omega |V(\omega,k_x)|, \tag{9}$$

$$k(f) \text{ peak:} \quad k_x(\omega) = \arg\max_k |V(\omega,k_x)|. \tag{10}$$

Equation 9 indicates that $f(k)$ peak can be computed from $dV/d\omega = 0$. Differentiation of Equation 8 with respect to $\omega$ followed by appropriate simplification results in,

$$\frac{dV}{d\omega} = 0 \quad \Rightarrow \quad \omega(k_x) = c_s k_x. \tag{11}$$

Similarly, the $k(f)$ peaks can be obtained from $dV/dk = 0$:

$$\frac{dV}{dk} = 0 \quad \Rightarrow \quad k_x(\omega) = \frac{\omega}{c_s\sqrt{1+\tau^2\omega^2}}. \tag{12}$$

The $f(k)$ and $k(f)$ peaks represented by Equations 11 and 12 are clearly different, except for the purely elastic case, i.e. $\tau = 0$. Conversely, the deviation between the two peaks is a function of $\tau$, indicating that $\tau$, thus the viscosity, can be estimated from the peak deviation. Elasticity can be obtained from the slope of the dispersion curve represented by Equation 12, which is the classical approach for estimating bulk tissue elasticity. We can thus conclude that both elasticity and viscosity can be estimated by matching both $f(k)$ and $k(f)$ peaks. Moreover, peaks are expected to be independent of the ARF magnitude and less sensitive to noise compared to the width of the peak.

Equations 11 and 12 can be used to match the experimental peaks in Figures 2b and 2c respectively but have several shortcomings: (a) the analytical peaks in Equations 11 and 12 are limited to the Kelvin-Voigt model, which may not be accurate for real tissues; (b) Equations 11 and 12 assume an idealized impulse for ARF, which may not be the case in reality; (c) diffraction correction through $\sqrt{r}$ modulation is not accurate in the near-field and in the presence of viscosity; and importantly, (d) signal processing details such as sampling and truncation in space and time affect the spread and thus the location of the experimental peaks but not the analytical peaks.

To overcome the above limitations, our approach to viscoelasticity estimation does not directly use Equations 11 and 12, but builds on the idea that the experimental peaks (Figure 2d) are sensitive to unknown viscoelasticity parameters. Specifically, we use inverse optimization approach where the particle velocity is obtained from a forward model on the $x-t$ grid that is identical to the experimental measurement. The simulated peaks are obtained from the simulated response using a procedure identical to that for computing the experimental peaks. The mismatch between the experimental and simulated peaks are quantified using least-squares error, which is iteratively minimized to estimate the viscoelastic parameters. In the remainder of the section, we provide the details of (a) the forward model to obtain the simulated response, (b) the approach to compute $f(k)$ and $k(f)$ peaks, followed by (c) the inversion approach used to solve for the unknown viscoelasticity parameters.

## Forward modeling for response computation

Given the homogeneous, unbounded approximation for the tissue, a Fourier transform is used to compute the response first in the wavenumber domain. Depending on the rheological models, the computation can either be done directly in the time domain, or in Fourier (frequency) domain. We present here both approaches, followed by a more efficient 2D approximation, which will make the inversion process more practical with respect to the computational effort.

### *Frequency-Wavenumber Approach for General Viscoelasticity*

Fourier transforming Equation 1 in both space and time results in,

$$-\rho\omega^2 U + \hat{G}k^2 U = F, \tag{13}$$

where $U = U(\boldsymbol{k},\omega)$ and $U = F(\boldsymbol{k},\omega)$ are the Fourier transforms of the displacement and ARF respectively. $\boldsymbol{k} = \{k_x, k_y, k_z\}^T$ is the wavenumber vector, where $k_x, k_y, k_z$ represent the spatial frequencies in $x, y, z$ directions. $k = \sqrt{k_x^2 + k_y^2 + k_z^2}$ is the magnitude of the wavenunber vector. Equation 13 immediately results in the particle displacement and velocity field:

$$U(\boldsymbol{k},\omega) = \frac{\hat{F}}{\hat{G}k^2 - \rho\omega^2}, \quad V(\boldsymbol{k},\omega) = \frac{i\omega\hat{F}}{\hat{G}k^2 - \rho\omega^2}. \tag{14}$$

Finally, inverse Fourier transforming in both space and time, results in the time-domain velocity response:

$$v(x,y,z,t) = IFT(V(\boldsymbol{k},\omega)). \tag{15}$$

Note that there are no restrictions on the underlying viscoelastic model and can consider Spring-Pot, Kelvin-voigt, Fractional Voigt, or a completely general viscoelastic model (along with appropriate parametrization needed for inversion).

### *Time-Wavenumber Approach for Kelvin-Voigt Model*

While the above frequency-wavenumber approach can be used with any viscoelastic model, time-wavenumber approach would be more efficient for the special case of Kelvin-Voigt model, and is presented here. For Kelvin-Voigt model the viscoelastic modulus can be written in an operator form as,

$$G = \rho c_s^2 \left(1 + \tau \frac{\partial}{\partial t}\right). \tag{16}$$

Substituting the above in Equation 1, and Fourier transforming in space results in,

$$\rho \frac{\partial^2 \bar{u}}{\partial t^2} + \rho c_s^2 k^2 \bar{u} + \rho c_s^2 k^2 \tau \frac{\partial}{\partial t} \bar{u} = \overline{F}, \tag{17}$$

where, $\bar{u} = \bar{u}(\boldsymbol{k},t)$ and $\overline{F} = \overline{F}(\boldsymbol{k},t)$ are the spatial Fourier transforms of the $u$ and $f$ respectively. The above equation essentially represents a damped vibration problem, i.e.,

$$M \frac{\partial^2 \bar{u}}{\partial t^2} + C \frac{\partial \bar{u}}{\partial t} + K\bar{u} = \overline{F}, \tag{18}$$

with mass $M = \rho$, damping $C = \rho c_s^2 k^2 \tau$ and stiffness $K = \rho c_s^2 k^2$. The natural frequency $\omega_n = \sqrt{K/M} = c_s k$ and the damping ratio, $\zeta = C/2\sqrt{KM} = c_s k \tau / 2$. Considering that ARF is a rectangular pulse in time (Equation 3), the resulting particle velocity can be derived as,

$$\bar{v}(\mathbf{k},t) = \frac{\bar{F}}{\rho(\lambda_1 - \lambda_2)} \left( e^{\lambda_1 t} - e^{\lambda_2 t} \right), \qquad t < t_d$$
$$= \frac{\bar{F}}{\rho(\lambda_1 - \lambda_2)} \left[ \left( e^{\lambda_1 t} - e^{\lambda_2 t} \right) - \left( e^{\lambda_1 (t-t_d)} - e^{\lambda_2 (t-t_d)} \right) \right], \qquad t > t_d \quad (19)$$

where,

$$\lambda_{1,2} = \omega_n ( -\zeta \pm \sqrt{\zeta^2 - 1} ). \quad (20)$$

We obtain the final space-time representation for the velocity response through inverse Fourier Transform in space:

$$v(x,y,z,t) = IFT(\bar{v}(\mathbf{k},t)) \quad (21)$$

*Two-Dimensional (2D) Approximation*

While the above subsections address 3D simulation in space, noting that the ARF push is often long in the $z$ direction and the response is often used only at the mid-depth, the simulation can be simplified from 3D to 2D. Essentially, the push, thus the response, are assumed to be independent of $z$ and vary only in the $x-y$ plane. This leads to 2D version of Equation 1 in the $x-y$ plane. When transformed into the Fourier domain, only wavenumbers $k_x$ and $k_y$ remain, with $k_z = 0$ (since no variation in $z$ direction). The remaining details of the formulation in the previous section stay the same. Such a simplification leads to significant savings in the computational cost (e.g. $5.5$ seconds for 2D analysis as opposed to $1250$ seconds for 3D, on a $12^{th}$ Gen Intel Core i9-12900k computer with $64$ GB RAM). These computational cost savings appear to come with minimal effects on accuracy as shown in the Results section. As illustrated in Figure 4b, 2D inversion of synthetic data generated from 3D forward model does not result in any significant errors, leading us to advocate the use of 2D forward models for TPM inversion.

## Identifying Peaks

The TPM is based on matching the experimental peaks with simulated peaks. Thus, the measured particle velocities from SWE data as well as the computed response from the forward modeling described above, must be converted to appropriate pair of $f(k)$ and $k(f)$ peaks. Care must be exercised in (a) reducing the effect of noise in the experimental data, and (b) avoiding any unintended discrepancies in the definition of peaks that could lead to errors in the eventual estimation of the viscoelasticity parameters.

Focusing first on the experimental peaks and associated noise effects, we start with the $x-t$ data as discretely sampled in the SWE motion data, with appropriate values of $\Delta x$, $\Delta t$, $x_{max}$ and $t_{max}$. A standard procedure in e.g. dispersion-based inversion is to correct for geometric spreading with $\sqrt{r}$ modulation. Based on our experience with experimental data, we do not use such modulation for TPM due to its effect on amplifying noise away from the ARF. More importantly, since we are using TPM by matching with the simulated peaks from a consistent forward model and not the simplified analytical peaks in Equations 11 and 12, such modulation is no longer necessary. At the end, the signal is appropriately padded and Fourier transformed in space and time to get the $f-k$ motion data. $f(k)$ peak is then the frequency $f$ corresponding to the maximum absolute $f-k$ response for any given wavenumber $k_x$. To visualize this peak, one can normalize absolute $f-k$ data for each $k_x$, by the maximum for that $k_x$, see e.g. Figure 2b. A similar procedure for can be followed for computing and visualizing $k(f)$ peaks (see Figure 2c).

For computing simulated peaks, it is important to keep in mind that many of the processing steps described for experimental data processing has effects on the peak locations, e.g. as highlighted by Rouze et al. [30], signal truncation in experimental data translates to convolution with the Sinc function in $f-k$ domain leading to increased spread and potential overestimation of viscoelasticity. We avoid such complexities by obtaining simulated peaks in the same way as the experimental peaks, i.e. the response obtained from the forward model is sampled and truncated exactly the same way as the SWE measurements, followed by using the same windowing and padding parameters to obtain the $f-k$ data that is used for obtaining the simulated peaks.

### Inversion using Twin Peaks

The unknown viscoelastic parameters of the tissues are estimated by matching simulated peaks with measured peaks. This is done by minimizing the objective function $F_{obj}$,

$$\mathbf{p}_{inv} = \arg\min_{\mathbf{p}} F_{obj}, \tag{22}$$

where $\mathbf{p}$ represents the vector of viscoelastic parameters and $\mathbf{p}_{inv}$ is the final estimate of the parameter vector. We define the objective function to be the sum of the relative least-squares differences between measured and simulated twin peaks:

$$F_{obj} = \sqrt{\frac{\sum_{i=1}^{N_1}\left(k_{s_i}^p - k_{m_i}^p\right)^2}{\sum_{i=1}^{N_1} k_{m_i}^{p^2}} + \frac{\sum_{i=1}^{N_2}\left(f_{s_i}^p - f_{m_i}^p\right)^2}{\sum_{i=1}^{N_2} f_{m_i}^{p^2}}}, \tag{23}$$

where $N_1$ is the number of points in $k(f)$ peaks, $N_2$ is the number of points in $f(k)$ peaks, $k_s^p$ and $f_s^p$ are simulated peaks, and $k_m^p$ and $f_m^p$ are measured peaks. The minimization in Equation 22 is carried out using the BFGS method, through MATLAB's "fminunc" function. Given the limited number of parameters, we utilize a finite difference derivative with a relative step size of $10^{-5}$.

*Choosing the frequency and wavenumber ranges.* The wavenumber and frequency ranges for peak matching is driven primarily by the data sampling parameters, i.e. $\Delta x, x_{\max}, \Delta t, t_{\max}$. Naturally, $\Delta x, \Delta t$ translate upper limits of the wavenumber and frequency ranges respectively, while $x_{\max}, t_{max}$ determine the lower limits of these ranges. In addition, to address the effects of noise in real data, the repeatability of the two peaks from across various replicates of SWE experiments is used to determine the ranges for reliable experimental peaks (this often has the effect of reducing the upper limits for viscous tissues, compared to more elastic tissues).

## Results

### Comparison between Frequency Domain and Time Domain Forward Models

In the forward model section, we introduced formulations to solve for the particle velocity in the frequency as well as the time domains. In this section, we briefly compare the results from both these forward models (2D versions), for the Kelvin-Voigt model with $G_0 = 2\,kPa$ and $\tau = 0.5\,ms$. For this analysis, we considered the measurement parameters to be, $\Delta x = 0.25\,mm$, $\Delta t = 0.1\,ms$, $x_{max} = 30\,mm$ and $t_{max} = 15\,ms$. The simulation is performed in both frequency and time domains, and the resulting peaks are presented in Figure 3a. Both $f(k)$ and $k(f)$ peaks match between the two analyses. The small deviation at the lower frequencies and wavenumbers result from not taking larger $x_{max}$ and $t_{max}$ for frequency domain simulation (the results for three-times larger values of $x_{max} = 90\,mm$ and $t_{max} = 45\,ms$ are shown in Figure 3b, confirming this hypothesis). Given the accuracy of time-domain simulation, combined with its efficiency (taking only $0.82$ second compared to $5.5$ seconds for frequency domain simulation), we advocate the use of time domain simulation whenever the Kelvin-Voigt model is used for inversion.

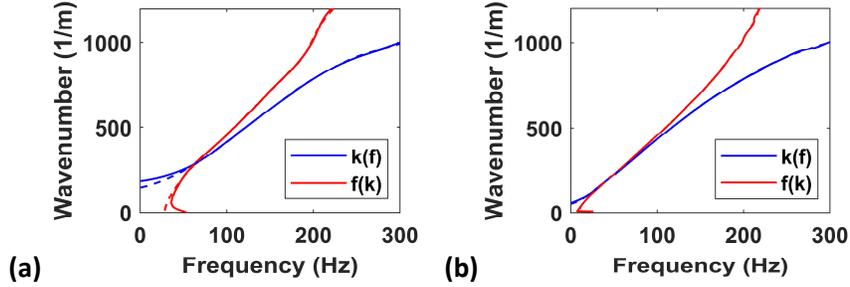

**Figure 3. Twin peaks from time domain simulation (dashed lines) and frequency domain (solid lines) simulation: (a) for $x_{max} = 30\,mm$ and $t_{max} = 15\,ms$; (b) for larger sampling window with $x_{max} = 90\,mm$ and $t_{max} = 45\,ms$.**

## *In silico* Verification

In this section, we perform *in silico* verification not only as a natural first step before validation with real experimental data, but also to further examine the accuracy of inversion based in 2D models when applied to *in silico* data obtained from 3D models, which is more representative of real experimental data. We perform such validation primarily for the Kelvin-Voigt model, followed by a brief exploration of the spring-pot model. We also compare the results from AMUSE method [28] and fw2DFT [30] in Figures 5 and 7.

### *Noise-Free Inversion with Kelvin-Voigt Model*

We start with a detailed inversion for the Kelvin-Voigt model with $G_0 = 2\,kPa$ and $\tau = 0.5\,ms$. The *in silico* data are obtained by detailed 3D simulation of wave propagation from ARF with $F/N = 1.25$ and attenuation $= 0.5\,dB/cm/MHz$. The ARF profile is computed using Field II, which is then used in the forward model using 3D time-wavenumber formulation [38,39]. The response at $z = 0$, i.e. at the center of the ARF, is used to generate the *in silico* peaks, which are then inverted using TPM with 2D forward model. The objective function defined in Equation 23 is plotted in Figure 4a, showing a bowl shape with a clear minimum. The match between *in silico* and inverted peaks is shown in Figure 4b, and the inverted parameters ($G_0 = 2.04\,kPa$ and $\tau = 0.51\,ms$) are close to the real parameters. These observations illustrate the robustness and accuracy of TPM, as well as the underlying 2D approximation in the forward model.

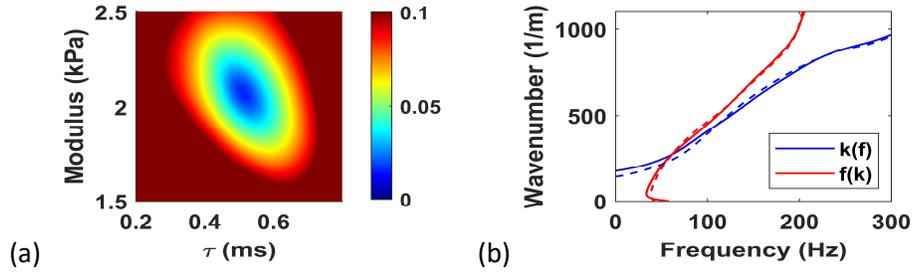

**Figure 4. (a) Objective function for noise-free *in silico* data, (b) twin peaks from the data (dashed) along with inverted peaks (solid)**

We then performed the verification with varying viscosity, i.e. with relaxation time ($\tau$) ranging from $0.1\,ms$ to $1.0\,ms$. For lower viscosity levels, i.e. $0.1 < \tau < 0.5\,ms$, peak-matching is performed for $100 < f < 500\,Hz$, $500 < k_x < 1500\,m^{-1}$. For higher viscosity, i.e. $0.5 < \tau < 1\,ms$, peak matching is performed for $100 < f < 200\,Hz$, $300 < k_x < 800\,m^{-1}$. Table 1 summarizes the inversion results, clearly illustrating the effectiveness of the TPM method even with a simpler 2D forward model.

Table 1. Inversion Results for Noise-free in-silico data

| Actual Parameter | | Inverted Parameter | | Error Percentage (%) | |
|---|---|---|---|---|---|
| $G_0(kPa)$ | $\tau(ms)$ | $G_0(kPa)$ | $\tau(ms)$ | $G_0$ | $\tau$ |
| 2.00 | 0.1 | 2.05 | 0.125 | 2.5 | 25.00 |
| 2.00 | 0.3 | 2.02 | 0.29 | 1.0 | 3.33 |
| 2.00 | 0.5 | 2.04 | 0.51 | 2.0 | 2.00 |
| 2.00 | 0.7 | 1.97 | 0.68 | 1.5 | 2.85 |
| 2.00 | 1.0 | 2.03 | 0.96 | 1.5 | 4.00 |

We also implemented the AMUSE method [28], and fw2DFT [30] procedure to estimate the viscoelasticity. As observed in Figure 5, the AMUSE method does not perform well at lower frequencies, for both elasticity (storage modulus) and viscosity (loss modulus). The fw2DFT method performs better than the AMUSE method at lower frequencies but has degraded performance at higher frequencies. The TPM, on the other hand, performs well across the frequency range. This is not surprising given the low number of parameters that are being inverted for the TPM; the other two methods do not assume any underlying rheological model, and could have improvements with an assumption of an underlying model, which is not explored here.

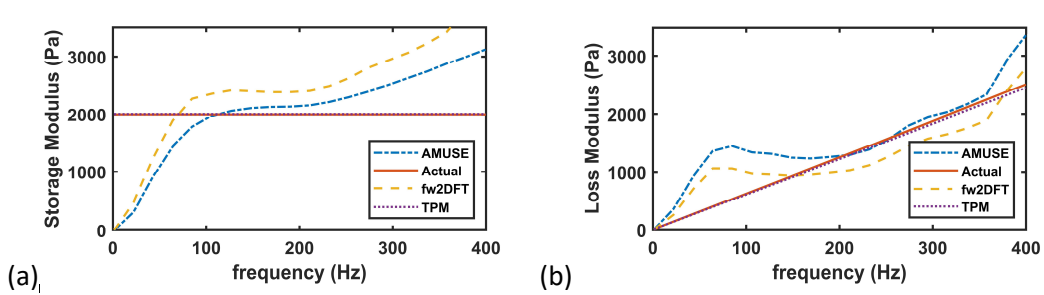

**Figure 5. Comparison of existing approaches with TPM: (a) storage modulus; (b) loss modulus.**

*Inversion with Noise-laden Synthetic Data*

We repeat the inversion in the previous section, now polluted with noise:

$$v_{noisy} = v_{synthetic}(1 + \alpha_m \varepsilon) + \alpha_a \varepsilon \times \max(v_{synthetic}), \tag{24}$$

where $\alpha_m$ and $\alpha_a$ denote multiplicative and additive noise levels and $\varepsilon$ is random signal generated from the standard normal distribution, i.e. with zero mean and unit standard deviation. We choose $\alpha_m = 0.30$ and $\alpha_a = 0.007$ for this study, informed by visual comparisons with real experimental data.

Figure 6a presents an example TPM objective function for the noisy data, which has a bowl shape with a clear minimum, indicating the robustness of TPM against noise. This observation on robustness is further reinforced by examining the match between the two peaks with the inverted peaks (see Figure 6b for an example, which is typical). The errors in inversion of $G_0$ is shown in Figure 6c, indicating $< 5\%$ error for various levels of relaxation time (the mean and standard deviations are shown for these errors with the thick lines representing the mean and the thin lines representing the standard deviation). The errors in viscosity are larger, at approximately $10\%$ up to relaxation time $0.45$ ms and approximately $5\%$ for higher relaxation times (shown in Figure 6d). As expected, these errors are larger than those from noise-free data but are still acceptable. Like for the noise-free data, we inverted the noisy data with AMUSE method as well as fw2DFT method. Since the AMUSE and fw2DFT methods are model-free, we post-processed the results to obtain the elastic modulus and relaxation time through least-squares fit of the storage and loss moduli in the considered frequency range. The final comparison is presented in Figure 7, leading to the conclusion that the TPM is able to provide more effective inversion of the viscoelastic parameters.

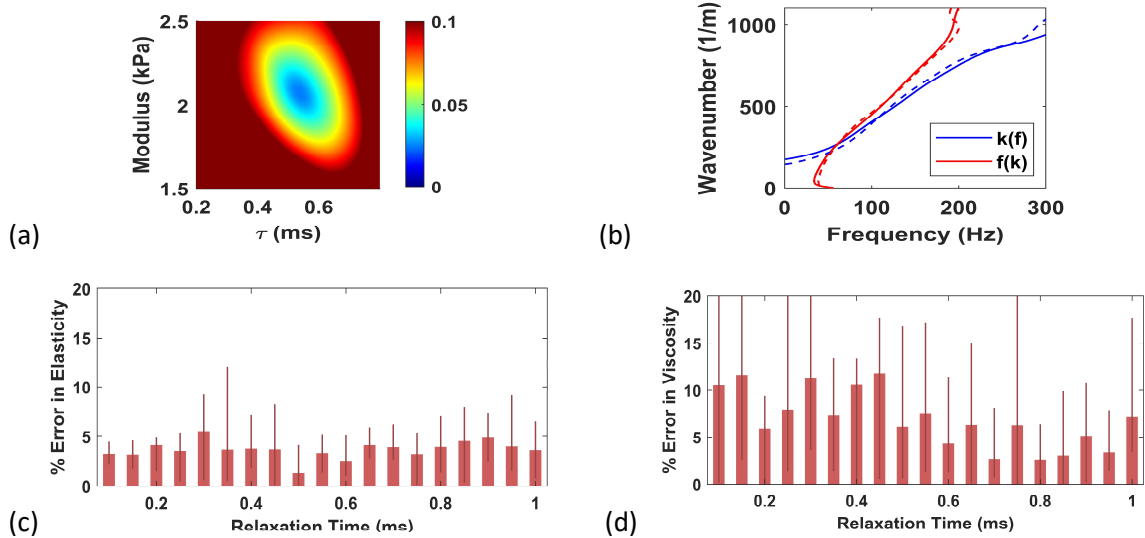

**Figure 6. (a) Objective function for noisy *in silico* data, (b) twin peaks from the data (dashed) and final inverted peaks (solid), (c) error in elastic modulus and d) error in viscosity for ten realizations.**

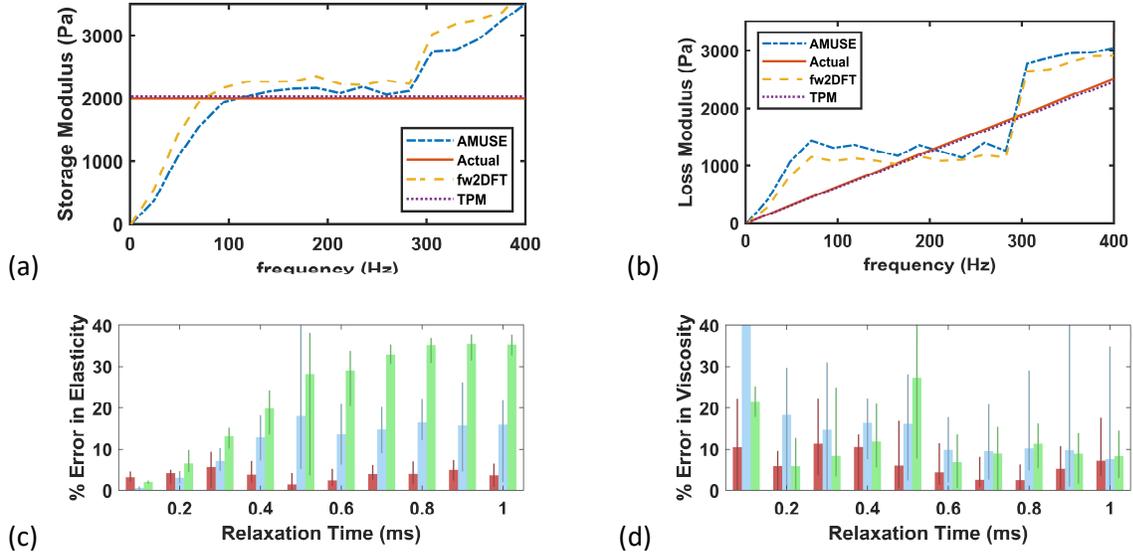

**Figure 7. Comparison of existing approaches with TPM: (a) example inverted storage modulus; (b) example inverted loss modulus; (c) error in elastic modulus; (d) error in viscosity (blue: AMUSE, green: fw2DFT and red: TPM).**

*Example Inversion with Spring-Pot Model*

In this section, we verified the TPM with the spring-pot viscoelastic model, i.e.,

$$G = G_0 \left( \frac{i\omega}{\omega_0} \right)^\alpha , \tag{25}$$

where $G_0$ is the modulus factor and $\alpha$ is the fractional order (note that $G_0$ in Equations 6 and 25 is not the same). The $\omega_0$ is the reference frequency introduced to maintain dimensional consistency and can be chosen arbitrarily. For this study, we chose $G_0 = 2000\ Pa$ and $\alpha = 0.4$ and $\omega_0 = 2\pi \times 200\ rad/s$. *In silico* data is obtained by detailed 3D simulation of wave propagation using the frequency domain forward model, followed by adding noise per Equation 24, again with $\alpha_m = 0.30$ and $\alpha_a = 0.007$, as informed by visual comparisons with real experimental data. The response at $z = 0$, i.e. at the center of the ARF, is used to generate the *in silico* peaks. These peaks are then inverted for $100\ Hz < f < 500\ Hz$ and $500\ m^{-1} < k_x < 1500\ m^{-1}$, using the 2D version of the forward model. The match between *in silico* and inverted peaks is shown in Figure 8a. The inverted parameters, $G_0 = 1975.2\ Pa$ and $\alpha = 0.4070$, are close to the real parameters with error of $<2\%$. The objective function defined in Equation 23 is plotted in Figure 8b which clearly illustrates a bowl shape with clear minimum, illustrating the robustness of the inversion.

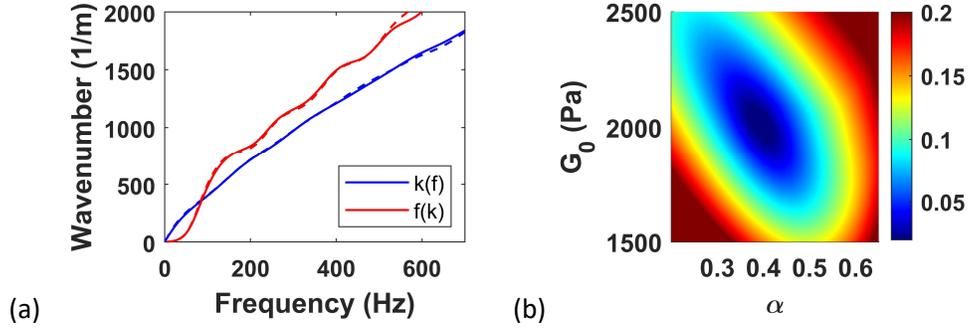

**Figure 8.** *(a)* Twin peaks obtained from noisy *in silico* data (dashed) and final inverted peaks (solid), *(b)* objective function.

## *Ex vivo* Validation

In this section, we validate the TPM with *ex vivo* testing on a porcine liver. SWE experiments were performed using a Verasonics V1 system (Verasonics, Inc., Kirkland, WA) equipped with a linear array transducer (L7-4, Philips Healthcare, Andover, MA). For the ARF push, an $F/N = 1.25$ was used with a $400\,\mu s$ at $4.09\,MHz$. The detection was performed using plane wave compounding $(-4°, 0°, +4°)$ with a pulse repetition period of $80\,\mu s$ between each plane wave transmission[40]. After compounding, the effective pulse repetition period was $240\,\mu s$ and an effective frame rate of $4166\,Hz$. The z component of the particle velocity for both *ex vivo* and *in vivo* resulting from ARF push was estimated from the acquired in-phase/quadrature (IQ) data using an autocorrelation method [41]. Six SWE experiments were performed at three apparently homogenous sites of the liver. A total of twelve datasets were generated considering left and right propagating waves. To reduce the noise, the signals are averaged over a 3 mm strip along the $z$ direction, centered at the focal depth of the ARF, resulting in the $x-t$ as illustrated in Figure 9a. The data is then truncated over $4 < x < 10\,mm$ and $0.535 < t < 10\,ms$ window as shown in Figure 9b, to highlight the main signal and eliminate the noise away from the signal. Examination of the resulting $x-t$ signals indicates that eleven out of the twelve datasets are suitable for the validation process.

For validation purposes, the liver was cored at four locations away from large blood vessels for independent characterization. The resulting $9.5\,mm$ cylindrical samples were mechanically tested with hyper-frequency viscoelastic spectroscopy ([42], Rheospectris C500+, Rheolution, Inc., Montreal, Quebec, Canada), to obtain the storage $(G_s)$ and loss $(G_l)$ moduli, which are shown in Figure 9d (dashed lines).

Out of eleven SWE responses, the peaks from three SWE measurements were clustered very closely together, while the scatter within the remaining peaks is more significant. Given this, we first inverted for these three clustered peaks together, which are shown as the dashed lines in Figure 9c. Inversion was performed for a frequency range of $0 < f < 200\,Hz$ and wavenumber range of $0 < k_x < 1200\,m^{-1}$. Informed by the Rheospectris measurements in Figure 9d, we chose the Kelvin-Voigt model for inversion, and the results are shown as the solid lines in Figure 9d. The inverted peaks are shown as solid lines in Figure 9c, illustrating a close match.

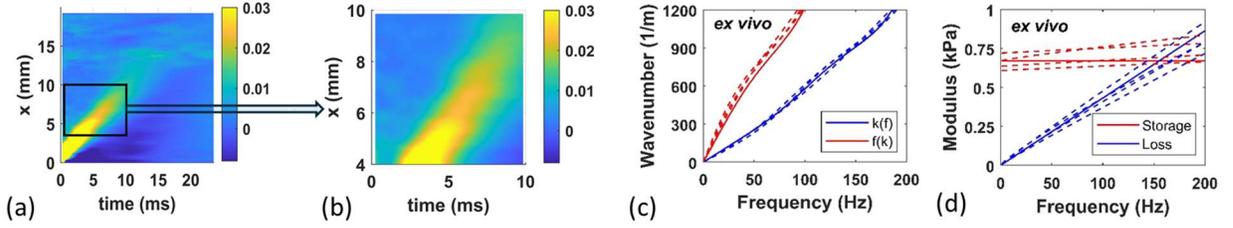

**Figure 9.** *Ex vivo* validation of TPM: (a) particle velocity data (for a single SWE sample), (b) truncated particle velocity data, (c) twin peaks for three SWE samples (dashed lines) along with inverted peaks (solid lines), (d) estimated moduli (solid lines) along with Rheospectris measurements (dashed lines)

We inverted the remaining eight SWE datasets together, using the same frequency and wavenumber ranges. The experimental (dashed lines) as well as inverted peaks (solid lines) are shown in Figure 10a. Figure 10b shows the comparison with Rheospectris data, indicating somewhat over prediction of viscosity. Noting the larger scatter in the $k(f)$ peaks at high frequencies, we refined the inversion by focusing on reduced frequency and wavenumber ranges, i.e. $0 < f < 200\ Hz, 0 < k_x < 1000\ m^{-1}$. The experimental peaks along with (modified) inverted peaks are shown in Figure 10c, illustrating reduced scatter in the peaks as well as better match with the inverted peaks. We also observed an improved match with the Rheospectris measurements (see Figure 10d). Based on these observations, it may be possible to utilize the repeatability of peaks as an indicator to automatically determine the frequency and wavenumber range that is best for inversion, which is left for future investigation. Notwithstanding such automation, the results presented indicate clear promise for using TPM for characterizing viscosity of real biological tissues.

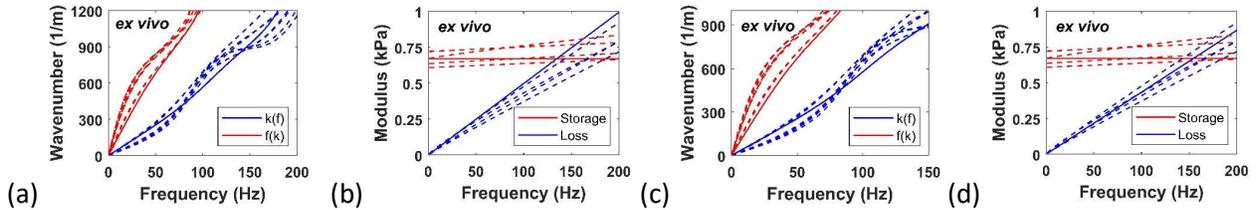

**Figure 10.** *Ex vivo* validation of TPM for remaining eight SWE samples: (a) twin peaks at higher $f-k$ range (dashed lines) along with inverted peaks (solid lines), (b) resulting inverted moduli (solid lines) with Rheospectris measurements (dashed lines); c) twin peaks at lower $f-k$ range; (d) resulting inverted moduli along with Rheospectris measurements.

### *In vivo* Application

To examine *in vivo* applicability of the TPM, we performed analysis on data from SWE experiments with a curved array transducer (C5-2v, Verasonics, Inc., Kirkland, WA) on *in vivo* human liver [43]. The data were collected based on a protocol approved by the Mayo Clinic Institutional Review Board and each subject provide written informed consent. Given the poor signal-to-noise ratio in this data (Figure 11a), the response is truncated further in the $x-t$ window as shown in Figure 11b for better characterization of SWE twin peaks (dashed lines in Figure 11c). We performed the inversion for frequency range of $60 < f < 170\ Hz$ and wavenumber range of $300 < k_x < 800\ m^{-1}$; these windows are narrower than *ex vivo* validation due to higher noise in the *in vivo* data. The inversion resulted in a storage shear modulus of *1.45 kPa* and a relaxation time of *0.63 ms*, which are in line with a healthy human liver [44,45]. While these numbers cannot be validated, the reliability and robustness of the inversion process can be examined through plotting the objective value $F_{obj}$ in Equation 23, as a function of the inversion parameters. This is shown in Figure 11d, which has a bowl shape, illustrating the robustness even in the presence of high noise typical in *in vivo* settings.

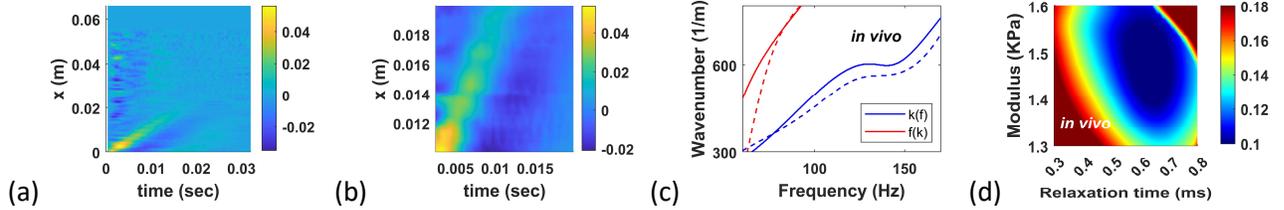

**Figure 11.** *In vivo* application: (a) unfiltered particle velocity; (b) truncated x-t data; (c) twin peaks (SWE is dashed line and simulated is solid line); (d) objective function.

## Summary and Conclusions

In an attempt to provide reliable point measurements of not just elasticity but viscoelasticity of soft tissues using shear wave elastography (SWE), we developed the so-called twin-peak method (TPM). The TPM is based on the observation that viscosity results in spreading of the particle velocity in the frequency-wavenumber $(f-k)$ domain, which in turn results in the separation of $f(k)$ peaks obtained by stepping through $k$, and $k(f)$ peaks obtained by stepping through $f$. Based on the premise that the peaks are typically less sensitive to noise compared to other amplitude-based measures, the TPM estimates the viscoelasticity parameters by matching these peaks from experimental measurements to those from simulation. The TPM is shown to be effective through verification using *in silico* data and validation using *ex vivo* data. The method is also applied to *in vivo* data, where the examination of the objective function lead to the observation that the TPM is robust and can be used even with *in vivo* data with significant noise.

While we largely focused on using Kelvin-Voigt model for viscoelasticity, the developed approach is applicable to general viscoelasticity, although the underlying model needs to be appropriately parameterized for the sake of inversion (we illustrated such an approach for the spring-pot model, while more complicated models can be considered depending on the context). The method can be further refined to automate the range of the frequencies and wavenumbers used for inversion. Finally, TPM provides point measurements, but it does so by matching the $f(k)$ and $k(f)$ peaks that inherently assume homogeneity within the measurement line (plane). This has an averaging effect and the associated approximation needs to be investigated when significant heterogeneities are involved; such cases may require imaging algorithms (see e.g. [46]), where the TPM results can potentially serve as initial estimates. Overall, TPM appears to provide a useful tool to measure tissue viscoelasticity that is accurate and robust against measurement noise.

## Acknowledgments

The work is partially funded by National Science Foundation grant DMS-2111234, and by National Institute of Health grant R01 HL145268. The content is solely the responsibility of authors and does not necessarily represent the official views of the National Science Foundation, the National Heart, Lung, and Blood Institute, or the National Institutes of Health.